\documentclass[prd,aps,nofootinbib,floatfix,11pt]{revtex4}

\usepackage{amsmath,graphicx,epsfig,amssymb,dsfont,mathtools}
\usepackage[usenames]{color}
\usepackage{ulem} %% for strike-through
\usepackage{bigstrut}
\usepackage{slashed}
\usepackage{multirow}
\usepackage{subfigure}

\begin{document}
	
\title{Weak decays of doubly heavy baryons: ``decay constants" }
	
\author{Xiao-Hui Hu$^{1}$,
%~\footnote{Email:huxiaohui@sjtu.edu.cn}, 
Yue-Long
Shen$^{2}$,
%~\footnote{Email:shenylmeteor@ouc.edu.cn}, 
Wei Wang$^{1}$~\footnote{Email:wei.wang@sjtu.edu.cn},
and Zhen-Xing Zhao$^{1}$~\footnote{Email:star\_0027@sjtu.edu.cn}}
	
\affiliation{$^{1}$ INPAC, Shanghai Key Laboratory for Particle Physics and Cosmology, \\
MOE Key Laboratory
for Particle Physics, Astrophysics and Cosmology,  \\
School of Physics and Astronomy, Shanghai Jiao-Tong University, Shanghai 200240, P.R.  China \\
$^2$ College of Information Science and Engineering, Ocean University of China,   Qingdao  266100, P.R. China }

\begin{abstract}
Inspired by the recent observation of the $\Xi_{cc}^{++}$ by LHCb collaboration, we explore the ``decay constants" of doubly heavy baryons in the framework of QCD sum rules. With the $\Xi_{cc}, \Xi_{bc}, \Xi_{bb}$, and  $\Omega_{cc}, \Omega_{bc}, \Omega_{bb}$ baryons interpolated by   three-quark operators, we calculate the correlation functions using the operator product expansion and include  the contribution from operators up to dimension six. On the hadron side, we consider both contributions from the lowest-lying states with $J^P=1/2^+$ and  from negative parity baryons with $J^P=1/2^-$. We find that the results are stable and  the  contaminations from negative parity baryons are not severe. These results are ingredients for the QCD study of weak decays and other properties of doubly-heavy baryons.  
\end{abstract}\maketitle

\section{Introduction}

%%%%%%%%%%%%%%%%%%%%%

It is widely believed that doubly heavy baryons with two charm and/or
bottom quarks exist in reality, but their experimental search has
been a while.   SELEX collaboration first reported the discovery
of $\Xi_{cc}^{+}$ in the $\Lambda_{c}^{+}K^{-}\pi^{+}$ final state
sixteen years ago \cite{Mattson:2002vu,Ocherashvili:2004hi}, with the 
mass measured as $m_{\Xi_{cc}^{+}}=(3519\pm1)$ MeV~\cite{Mattson:2002vu,Ocherashvili:2004hi}.
However, the SELEX-like $\Xi_{cc}^{+}$ signal is not confirmed by
later experiments \cite{Ratti:2003ez,Aubert:2006qw,Chistov:2006zj,Aaij:2013voa,Kato:2013ynr}.
In 2017, in the $\Lambda_{c}^{+}K^{-}\pi^{+}\pi^{+}$ final state
the LHCb collaboration has observed the doubly charmed baryon $\Xi_{cc}^{++}$
with the mass~\cite{Aaij:2017ueg}: 
\begin{equation}
m_{\Xi_{cc}^{++}}=(3621.40\pm0.72\pm0.27\pm0.14)\ {\rm MeV}.\label{eq:LHCb_measurement}
\end{equation}
In order to decipher the internal nature of doubly heavy baryons and
uncover the underlying dynamics in the transition, more experimental
investigations of the production and decays are  heavily  demanded.
Meanwhile further theoretical studies on weak decays of doubly heavy
baryons will be of great importance~\cite{Chen:2017sbg,Yu:2017zst,Wang:2017mqp,Li:2017cfz,Meng:2017udf,Wang:2017azm,Karliner:2017qjm,Gutsche:2017hux,Li:2017pxa,Guo:2017vcf,Lu:2017meb,Xiao:2017udy,Sharma:2017txj,Ma:2017nik,Meng:2017dni,Li:2017ndo,Wang:2017qvg,Shi:2017dto,Xiao:2017dly,Yao:2018zze,Zhao:2018mrg,Xing:2018lre},
and in particular the solid QCD analyses of weak decays and production
are heavily requested.

In this work, we will present an analysis of the ``decay constant\char`\"{}
of doubly heavy baryons in the framework of QCD sum rules (QCDSR).  QCDSR has been extensively applied to study the hadron masses,
decay constants and transition form factors, the mixing matrix elements
of $K$-meson and $B$-meson systems, etc.~\cite{Colangelo:2000dp,Reinders:1986gv,Castillo:2003pt,Kiselev:1988it,Narison:1987qc,Bilic:1987gk,Reinders:1988aa,Penin:2001ux,Neubert:1991sp,Colangelo:1992cx}.
In this approach, hadrons are interpolated by the corresponding quark
operators. The correlation function of these operators can be handled
using the operator product expansion (OPE), where the short-distance
coefficients and long-distance quark-gluon interactions are separated.
The former are calculable  in  QCD perturbation theory, whereas the
latter can be parameterized in terms of vacuum condensates. The QCD result is then matched, via dispersion relation, onto  the observable characteristics of hadronic states. Due to
various advantages, the QCDSR has been used to calculate the masses
of doubly heavy baryons in Refs.~\cite{Zhang:2008rt,Wang:2010hs,Chen:2017sbg,Wang:2017qvg,Wang:2010it,Wang:2010vn,Aliev:2012ru,Aliev:2012iv}.
The main motif of this work is to study ``decay constants\char`\"{}
using the QCDSR. The ``decay constants\char`\"{} defined by the interpolating
current are mandatory inputs for studies of other properties of doubly
heavy baryons in QCDSR, for example the heavy-to-light transition
form factors.

The rest of the paper is arranged as follows. In Sec.~II, we will
present the calculation of correlation function in QCD sum rules,
including the explicit expressions of the spectral functions. We include
both the contributions from the $J^{P}=1/2^{+}$ baryons and the contamination
from the $J^{P}=1/2^{-}$ baryons. Sec.~III is devoted to the numerical
results. A summary is presented in the last section.

%%%%%%%%%%%%%%%%%%%%%%%%%%	

\section{QCD sum rules study }

%%%%%%%%%%%%%%%%%%%%%%%%%%

%%%%%%%%%%%%%%%%%%%%%%%%%%%%
\begin{table*}[!htb]
\caption{Quantum numbers and quark content for the ground state of doubly heavy
baryons. The $s_{h}$ denotes the spin of the heavy quark system. }
\label{tab:JPC} %
\begin{tabular}{cccc|cccc}
\hline 
{ Baryon }  & { Quark Content }  & { $s_{h}^{\pi}$ }  & { $J^{P}$ }  & { Baryon }  & { Quark Content }  & { $s_{h}^{\pi}$ }  & { $J^{P}$ }\tabularnewline
\hline 
{ $\Xi_{cc}$ }  & { $\{cc\}q$ }  & { $1^{+}$ }  & { $1/2^{+}$ }  & { $\Xi_{bb}$ }  & { $\{bb\}q$ }  & { $1^{+}$ }  & { $1/2^{+}$ }\tabularnewline
{ $\Xi_{cc}^{*}$ }  & { $\{cc\}q$ }  & { $1^{+}$ }  & { $3/2^{+}$ }  & { $\Xi_{bb}^{*}$ }  & { $\{bb\}q$ }  & { $1^{+}$ }  & { $3/2^{+}$ }\tabularnewline
\hline 
{ $\Omega_{cc}$ }  & { $\{cc\}s$ }  & { $1^{+}$ }  & { $1/2^{+}$ }  & { $\Omega_{bb}$ }  & { $\{bb\}s$ }  & { $1^{+}$ }  & { $1/2^{+}$ }\tabularnewline
{ $\Omega_{cc}^{*}$ }  & { $\{cc\}s$ }  & { $1^{+}$ }  & { $3/2^{+}$ }  & { $\Omega_{bb}^{*}$ }  & { $\{bb\}s$ }  & { $1^{+}$ }  & { $3/2^{+}$ }\tabularnewline
\hline 
{ $\Xi_{bc}'$ }  & { $\{bc\}q$ }  & { $0^{+}$ }  & { $1/2^{+}$ }  & { $\Omega_{bc}'$ }  & { $\{bc\}s$ }  & { $0^{+}$ }  & { $1/2^{+}$ }\tabularnewline
{ $\Xi_{bc}$ }  & { $\{bc\}q$ }  & { $1^{+}$ }  & { $1/2^{+}$ }  & { $\Omega_{bc}$ }  & { $\{bc\}s$ }  & { $1^{+}$ }  & { $1/2^{+}$ }\tabularnewline
{ $\Xi_{bc}^{*}$ }  & { $\{bc\}q$ }  & { $1^{+}$ }  & { $3/2^{+}$ }  & { $\Omega_{bc}^{*}$ }  & { $\{bc\}s$ }  & { $1^{+}$ }  & { $3/2^{+}$ }\tabularnewline
\hline 
\end{tabular}
\end{table*}

%%%%%%%%%%%%%%%%%%%%%%%%%%%%

A doubly heavy baryon is made of two heavy quarks and one light quark.
The quantum numbers and quark contents for the ground states are given
in Table \ref{tab:JPC}. In this work we will study the $J^{P}=1/2^{+}$
baryons which can only weakly decay.

\subsection{QCD Sum rules with only positive parity baryons}

The interpolating current for the $\Xi_{QQ}$ and $\Omega_{QQ}$ is
chosen as 
\begin{eqnarray}
J_{\Xi_{QQ}} & = & \epsilon_{abc}\left(Q_{a}^{T}C\gamma^{\mu}Q_{b}\right)\gamma_{\mu}\gamma_{5}q_{c},\label{eq:current_Xi_QQ}\\
J_{\Omega_{QQ}} & = & \epsilon_{abc}\left(Q_{a}^{T}C\gamma^{\mu}Q_{b}\right)\gamma_{\mu}\gamma_{5}s_{c},
\end{eqnarray}
where $Q=c$ or $Q=b$. For the $\Xi_{bc}$ and $\Omega_{bc}$, we
choose 
\begin{eqnarray}
J_{\Xi_{bc}} & = & \frac{1}{\sqrt{2}}\epsilon_{abc}\left(b_{a}^{T}C\gamma^{\mu}c_{b}+c_{a}^{T}C\gamma^{\mu}b_{b}\right)\gamma_{\mu}\gamma_{5}q_{c},\\
J_{\Omega_{bc}} & = & \frac{1}{\sqrt{2}}\epsilon_{abc}\left(b_{a}^{T}C\gamma^{\mu}c_{b}+c_{a}^{T}C\gamma^{\mu}b_{b}\right)\gamma_{\mu}\gamma_{5}s_{c}.\label{eq:current_Omega_bc}
\end{eqnarray}
In the above equations, we have considered the $s_{h}^{\pi}=1^{+}$
baryons only.

The QCDSR analysis starts with the two-point correlator: 
\begin{eqnarray}
\Pi(q) & = & i\int d^{4}xe^{iq\cdot x}\langle0|T[J(x),{\bar{J}}(0)]|0\rangle,\label{eq:correlation_function}
\end{eqnarray}
where the interpolating current has been given in the above, and $\overline{J}$
is defined as 
\begin{eqnarray}
\bar{J}=J^{\dagger}\gamma^{0}.
\end{eqnarray}
A Lorentz structure analysis implies that the two-point correlation
function has the form: 
\begin{equation}
\Pi(q)=q\!\!\!\slash\Pi_{1}(q^{2})+\Pi_{2}(q^{2}).
\end{equation}
On the hadronic side, one can insert the complete set of hadronic
states into the correlator and then the correlator can be expressed
as a dispersion integral over a physical spectral function: 
\begin{eqnarray}
\Pi(q) & = & \lambda_{H}^{2}\frac{q\!\!\!\slash+m_{H}}{m_{H}^{2}-q^{2}}+\frac{1}{\pi}\int_{s_{0}}^{\infty}ds\frac{{\rm Im}\Pi(s)}{s-q^{2}},\label{eq:hadronic_level}
\end{eqnarray}
where $H$ can be a ground-state doubly heavy baryon and $m_{H}$
denotes its mass. In obtaining the above expression, the polarization
summation for spinors has been used: 
\begin{equation}
\sum_{s}u(q,s)\bar{u}(q,s)=q\!\!\!\slash+m_{H}.
\end{equation}
The pole residue $\lambda_{H}$ is defined as 
\begin{equation}
\langle0|J_{H}|H(q,s)\rangle=\lambda_{H}u(q,s).\label{eq:decay_constant_1/2+}
\end{equation}
The mass dimension for $\lambda_{H}$ is 3, while in analogy with
the meson case, it is convenient to use the ``decay constant\char`\"{}
with the definition 
\begin{equation}
\langle0|J_{H}|H(q,s)\rangle=f_{H}m_{H}^{2}u(q,s).
\end{equation}

In the OPE side, we will work at leading order in $\alpha_{s}$ in
this work and include the condensate contributions up to dimension
six. The full propagator for the heavy quark is given as 
\begin{eqnarray}
S_{ij}^{Q}(x) & = & i\int\frac{d^{4}k}{(2\pi)^{4}}e^{-ik\cdot x}\bigg[\frac{\delta_{ij}}{k\!\!\!\slash-m_{Q}}-\frac{g_{s}G_{\alpha\beta}^{a}t_{ij}^{a}}{4}\frac{\sigma^{\alpha\beta}(k\!\!\!\slash+m_{Q})+(k\!\!\!\slash+m_{Q})\sigma^{\alpha\beta}}{(k^{2}-m_{Q}^{2})^{2}}\nonumber \\
 &  & +\frac{g_{s}D_{\alpha}G_{\beta\lambda}^{n}t_{ij}^{n}(f^{\lambda\beta\alpha}+f^{\lambda\alpha\beta})}{3(k^{2}-m_{Q}^{2})^{4}}-\frac{g_{s}^{2}(t^{a}t^{b})_{ij}G_{\alpha\beta}^{a}G_{\mu\nu}^{b}(f^{\alpha\beta\mu\nu}+f^{\alpha\mu\beta\nu}+f^{\alpha\mu\nu\beta})}{4(k^{2}-m_{Q}^{2})^{5}}\bigg],
\end{eqnarray}
with 
\begin{eqnarray}
f^{\lambda\alpha\beta} & = & (k\!\!\!\slash+m_{Q})\gamma^{\lambda}(k\!\!\!\slash+m_{Q})\gamma^{\alpha}(k\!\!\!\slash+m_{Q})\gamma^{\beta}(k\!\!\!\slash+m_{Q}),\\
f^{\alpha\beta\mu\nu} & = & (k\!\!\!\slash+m_{Q})\gamma^{\alpha}(k\!\!\!\slash+m_{Q})\gamma^{\beta}(k\!\!\!\slash+m_{Q})\gamma^{\mu}(k\!\!\!\slash+m_{Q})\gamma^{\nu}(k\!\!\!\slash+m_{Q}),
\end{eqnarray}
where $t^{n}=\lambda^{n}/2$ and $\lambda^{n}$ is the Gell-Mann matrix,
and the $i,j$ are the color indices. The full propagator for light
quarks is given as 
\begin{eqnarray}
S_{ij}(x) & = & \frac{i\delta_{ij}x\!\!\!\slash}{2\pi^{2}x^{4}}-\frac{\delta_{ij}}{12}\langle\bar{q}q\rangle-\frac{\delta_{ij}x^{2}\langle\bar{q}g_{s}\sigma Gq\rangle}{192}+\frac{i\delta_{ij}x^{2}x\!\!\!\slash\langle\bar{s}g_{s}\sigma Gs\rangle m_{q}}{1152}\nonumber \\
 &  & -\frac{ig_{s}G_{\alpha\beta}t_{ij}^{a}(x\!\!\!\slash\sigma^{\alpha\beta}+\sigma^{\alpha\beta}x\!\!\!\slash)}{32\pi^{2}x^{2}}.
\end{eqnarray}
With the quark propagators one can express the correlation function
in terms of a dispersion relation as: 
\begin{eqnarray}
\Pi_{i}(q^{2}) & = & \int_{(m_{Q}+m_{Q'})^{2}}^{\infty}ds\frac{\rho_{i}(s)}{s-q^{2}},\;\;\;i=1,2,
\end{eqnarray}
where the spectral density is given by the imaginary part of the correlation
function: 
\begin{eqnarray}
\rho_{i}(s) & = & \frac{1}{\pi}{\rm Im}\Pi_{i}^{{\rm OPE}}(s).
\end{eqnarray}
After equating the two expressions for $\Pi(q^{2})$ based on the
quark-hadron duality, and making a Borel transformation, we can write
the sum rules as 
\begin{eqnarray}
\lambda_{H}^{2}e^{-m_{H}^{2}/M^{2}} & = & \int_{(m_{Q}+m_{Q'})^{2}}^{s_{0}}ds\rho_{1}(s)e^{-s/M^{2}},\label{eq:sum_rule_1}\\
\lambda_{H}^{2}m_{H}e^{-m_{H}^{2}/M^{2}} & = & \int_{(m_{Q}+m_{Q'})^{2}}^{s_{0}}ds\rho_{2}(s)e^{-s/M^{2}}.\label{eq:sum_rule_1_no_use}
\end{eqnarray}
The spectral functions $\rho_{1}$ and $\rho_{2}$ are given as follows:
\begin{eqnarray}
\rho_{1}^{{\rm pert}}(s) & = & \frac{6}{(2\pi)^{4}}\int_{\alpha_{\min}}^{\alpha_{\max}}\frac{d\alpha}{\alpha}\int_{\beta_{\min}}^{1-\alpha}\frac{d\beta}{\beta}\bigg([\alpha\beta s-\alpha m_{Q}^{2}-\beta m_{Q^{\prime}}^{2}]^{2}\nonumber \\
 &  & \;\;\;\;+(1-\alpha-\beta)m_{Q}m_{Q^{\prime}}[\alpha\beta s-\alpha m_{Q}^{2}-\beta m_{Q^{\prime}}^{2}]\bigg),\\
\rho_{1}(s) & = & \rho_{1}^{{\rm pert}}(s)+\frac{\langle g_{s}^{2}G^{2}\rangle}{72}\left(m_{Q}^{2}\frac{\partial^{3}}{(\partial m_{Q}^{2})^{3}}+m_{Q^{\prime}}^{2}\frac{\partial^{3}}{(\partial m_{Q^{\prime}}^{2})^{3}}\right)\rho_{1}^{{\rm pert}}(s)\nonumber \\
 &  & +\frac{4m_{Q}m_{Q^{\prime}}\langle g_{s}^{2}G^{2}\rangle}{(4\pi)^{4}}\bigg(\frac{\partial^{2}}{(\partial m_{Q}^{2})^{2}}+\frac{\partial^{2}}{(\partial m_{Q^{\prime}}^{2})^{2}}\bigg)\int_{\alpha_{{\rm min}}}^{\alpha_{{\rm max}}}\frac{d\alpha}{\alpha}\int_{\beta_{{\rm min}}}^{1-\alpha}\frac{d\beta}{\beta}(1-\alpha-\beta)(\alpha\beta s-\alpha m_{Q}^{2}-\beta m_{Q^{\prime}}^{2})\nonumber \\
 &  & +\frac{2\langle g_{s}^{2}G^{2}\rangle}{(4\pi)^{4}}\bigg(\frac{\partial}{\partial m_{Q}^{2}}+\frac{\partial}{\partial m_{Q^{\prime}}^{2}}\bigg)\int_{\alpha_{{\rm min}}}^{\alpha_{{\rm max}}}d\alpha\int_{\beta_{{\rm min}}}^{1-\alpha}d\beta(3\alpha m_{Q}^{2}+3\beta m_{Q^{\prime}}^{2}-m_{Q}m_{Q^{\prime}}-4\alpha\beta s)\label{eq:rho1}\\
\rho_{2}(s) & = & -\frac{\langle\bar{q}q\rangle}{2\pi^{2}}\int_{\alpha_{\min}}^{\alpha_{\max}}d\alpha(3\alpha(1-\alpha)s-2\alpha m_{Q}^{2}-2(1-\alpha)m_{Q^{\prime}}^{2}+2m_{Q}m_{Q^{\prime}})\nonumber \\
 &  & -\frac{\langle\bar{q}g_{s}\sigma Gq\rangle}{8\pi^{2}}\left(1+\frac{s}{M^{2}}\right)A(s)-\frac{2\langle\bar{q}g_{s}\sigma Gq\rangle}{(4\pi)^{2}}\bigg((\alpha_{\max}-\alpha_{\min})\nonumber \\
 &  & +\frac{1}{2s(\alpha_{\max}-\alpha_{\min})}[\alpha_{\max}(1-\alpha_{\max})s+\alpha_{\min}(1-\alpha_{\min})s+4m_{Q}m_{Q^{\prime}}]\bigg),\label{eq:rho2}
\end{eqnarray}
with 
\begin{equation}
A(s)=\frac{-s^{3}+(m_{Q}^{2}+m_{Q^{\prime}}^{2})s^{2}+(m_{Q}^{2}-4m_{Q}m_{Q^{\prime}}+m_{Q^{\prime}}^{2})[s(m_{Q}^{2}+m_{Q^{\prime}}^{2})-(m_{Q}^{2}-m_{Q^{\prime}}^{2})^{2}]}{2s^{2}\sqrt{(s+m_{Q}^{2}-m_{Q^{\prime}}^{2})^{2}-4m_{Q}^{2}s}}.
\end{equation}
The integration limits are given by $\alpha_{\min}=[s-m_{Q}^{2}+m_{Q^{\prime}}^{2}-\sqrt{(s-m_{Q}^{2}+m_{Q^{\prime}}^{2})^{2}-4m_{Q^{\prime}}^{2}s}]/(2s)$,
$\alpha_{\max}=[s-m_{Q}^{2}+m_{Q^{\prime}}^{2}+\sqrt{(s-m_{Q}^{2}+m_{Q^{\prime}}^{2})^{2}-4m_{Q^{\prime}}^{2}s}]/(2s)$,
and $\beta_{\min}=\alpha m_{Q}^{2}/(s\alpha-m_{Q^{\prime}}^{2})$.
For the $\Omega_{QQ'}$, one needs to replace the condensates correspondingly.
The integration lower bound $(m_{Q}+m_{Q'})^{2}$ is replaced by $(m_{Q}+m_{Q'}+m_{s})^{2}$.

In Ref.~\cite{Zhang:2008rt}, the authors obtained a similar expression
with our Eq.~(\ref{eq:sum_rule_1}): 
\begin{eqnarray}
\rho_{1}(s) & = & -\frac{3}{2^{4}\pi^{4}}\int_{\alpha_{{\rm min}}}^{\alpha_{{\rm max}}}\frac{d\alpha}{\alpha}\int_{\beta_{{\rm min}}}^{1-\alpha}\frac{d\beta}{\beta}[\alpha\beta s-\alpha m_{Q}^{2}-\beta m_{Q'}^{2}]^{2}\nonumber \\
 &  & +\frac{3}{2^{2}\pi^{4}}m_{Q}m_{Q'}\int_{\alpha_{{\rm min}}}^{\alpha_{{\rm max}}}\frac{d\alpha}{\alpha}\int_{\beta_{{\rm min}}}^{1-\alpha}\frac{d\beta}{\beta}(1-\alpha-\beta)[\alpha\beta s-\alpha m_{Q}^{2}-\beta m_{Q'}^{2}]\nonumber \\
 &  & -\frac{5m_{q}\langle\bar{q}q\rangle}{2^{3}\pi^{2}}\int_{\alpha_{{\rm min}}}^{\alpha_{{\rm max}}}d\alpha\alpha(1-\alpha).\label{eq:rho1_huang}
\end{eqnarray}
A few remarks are in order. 
\begin{itemize}
\item We did not include the mass corrected quark condensate. This might
has some impact in the case of $\Omega_{cc,bc,bb}$.
\item However the gluon condensate contribution, which is anticipated more
important, is missing in Eq.~(\ref{eq:rho1_huang}). 
\item It should be noted that in the massless limit, we have the spectral
function: 
\begin{equation}
\rho_{1}(s)=\frac{s^{2}}{64\pi^{4}}+\frac{2\langle g_{s}^{2}G^{2}\rangle}{(4\pi)^{4}}.\label{eq:rho1_limit}
\end{equation}
Our result is fully consistent with Ref.~\cite{Braun:2000kw}: 
\begin{equation}
\lambda_{H}^{2}e^{-m_{H}^{2}/M^{2}}=\frac{1}{2(2\pi)^{4}}\left[M^{6}\left(1-e^{-s_{0}/M^{2}}\left(1+\frac{s_{0}}{M^{2}}+\frac{1}{2}\frac{s_{0}^{2}}{M^{4}}\right)\right)+\frac{\langle g_{s}^{2}G^{2}\rangle}{4}M^{2}\left(1-e^{-s_{0}/M^{2}}\right)\right].
\end{equation}
\item In Ref.~\cite{Zhang:2008rt}, the predicted mass $m_{\Xi_{cc}}=(4.26\pm0.19)\ {\rm GeV}$
is much larger than the experimental data $m_{\Xi_{cc}^{++}}^{{\rm exp}}=3.621\,{\rm GeV}$.
\end{itemize}

\subsection{QCD Sum rules with both positive and negative parity baryons}

In the above analysis, only the $1/2^{+}$ baryons are considered.
An interpolating current for the negative parity $1/2^{-}$ baryon
can be defined as 
\begin{eqnarray}
J_{-}\equiv i\gamma_{5}J_{+},
\end{eqnarray}
where $J_{+}$ is given in Eqs.~(\ref{eq:current_Xi_QQ}-\ref{eq:current_Omega_bc}).
When the complete set of hadron states is inserted to the correlation
function in Eq.~\eqref{eq:correlation_function}, both the positive
and the negative parity single-particle states can contribute~\cite{Jido:1996ia,Khodjamirian:2011jp}.

When taking into account the $1/2^{-}$ single-particle states, Eq.
(\ref{eq:hadronic_level}) is rewritten as 
\begin{eqnarray}
\Pi(q) & = & \lambda_{+}^{2}\frac{q\!\!\!\slash+m_{+}}{m_{+}^{2}-q^{2}}+\lambda_{-}^{2}\frac{q\!\!\!\slash-m_{-}}{m_{-}^{2}-q^{2}}+\frac{1}{\pi}\int_{s_{0}}^{\infty}ds\frac{{\rm Im}\Pi(s)}{s-q^{2}},\label{eq:hadronic_level_2}
\end{eqnarray}
where $\lambda_{\pm}$ ($m_{\pm}$) stands for the ``decay constant\char`\"{}
(mass) of positive or negative parity baryons. Apparently, the $\lambda_{+}$
is the ``decay constant\char`\"{} $\lambda_{H}$ we have defined
in Eq.~\eqref{eq:decay_constant_1/2+}. The $\lambda_{-}$ is defined
as 
\begin{equation}
\langle0|J_{H}^{+}|H(1/2^{-},q,s)\rangle=i\gamma_{5}\lambda_{-}u(q,s).
\end{equation}

At the hadronic level, one can take the imaginary part of the correlation
function as follows: 
\begin{eqnarray}
\frac{1}{\pi}{\rm Im}\Pi(s) & = & \lambda_{+}^{2}(\slashed q+m_{+})\delta(s-m_{+}^{2})+\lambda_{-}^{2}(\slashed q-m_{-})\delta(s-m_{-}^{2})+\cdots\nonumber \\
 & = & \slashed q\rho_{1}^{{\rm had}}(s)+\rho_{2}^{{\rm had}}(s),
\end{eqnarray}
with 
\begin{eqnarray}
\rho_{1}^{{\rm had}}(s) & = & \lambda_{+}^{2}\delta(s-m_{+}^{2})+\lambda_{-}^{2}\delta(s-m_{-}^{2})]+\cdots,\nonumber \\
\rho_{2}^{{\rm had}}(s) & = & m_{+}\lambda_{+}^{2}\delta(s-m_{+}^{2})-m_{-}\lambda_{-}^{2}\delta(s-m_{-}^{2})]+\cdots.
\end{eqnarray}
Here the ellipses stand for the contributions from higher resonances
and the continuum spectra. Considering the combination $\sqrt{s}\rho_{1}^{{\rm had}}+\rho_{2}^{{\rm had}}$,
and introducing the exponential function $\exp(-s/M^{2})$ to suppress
these contributions, one can separate the $\lambda_{+}$ contributions:
\begin{equation}
\int_{\Delta}^{s_{0}}ds[\sqrt{s}\rho_{1}^{{\rm had}}(s)+\rho_{2}^{{\rm had}}(s)]\exp\left(-\frac{s}{M^{2}}\right)=2m_{+}\lambda_{+}^{2}\exp\left(-\frac{m_{+}^{2}}{M^{2}}\right),
\end{equation}
where $s_{0}$ is the threshold of the continuum states and $M^{2}$
is the Borel parameter.

On the OPE side, we compute the correlation function $\Pi(q)$ to
obtain the QCD spectral densities 
\begin{equation}
\frac{1}{\pi}{\rm Im}\Pi(s)=\slashed q\rho_{1}^{{\rm OPE}}(s)+\rho_{2}^{{\rm OPE}}(s).
\end{equation}

Taking the quark-hadron duality below the continuum threshold $s_{0}$,
we arrive at the following QCD sum rule 
\begin{equation}
2m_{+}\lambda_{+}^{2}\exp\left(-\frac{m_{+}^{2}}{M^{2}}\right)=\int_{\Delta}^{s_{0}}ds[\sqrt{s}\rho_{1}^{{\rm OPE}}(s)+\rho_{2}^{{\rm OPE}}(s)]\exp\left(-\frac{s}{M^{2}}\right).\label{eq:sum_rule_2}
\end{equation}
Here $\Delta$ is the threshold parameter, $\Delta=(m_{Q}+m_{Q^{\prime}})^{2}$
for $\Xi_{QQ^{\prime}}$, and $\Delta=(m_{Q}+m_{Q^{\prime}}+m_{s})^{2}$
for $\Omega_{QQ^{\prime}}$.

\section{Numerical Results}

In the numerical analysis, the quark masses are used as~\cite{Olive:2016xmw}:
$m_{c}=1.35\pm0.10\;{\rm GeV},\;m_{b}=4.60\pm0.10~{\rm GeV},\;m_{s}=0.12\pm0.01\,{\rm GeV}$,
while the $u$ and $d$ quarks are taken as massless. Similar values
have been taken in Ref.~\cite{Wang:2010hs}.

The vacuum condensates are used as \cite{Shifman:1978bx,Reinders:1984sr,Colangelo:2000dp,Colangelo:1995qp,Zhang:2008rt,Wang:2016mee}:
$\langle\bar{q}q\rangle=-(0.24\pm0.01\,{\rm GeV})^{3}$, $\langle\bar{s}s\rangle=(0.8\pm0.1)\langle\bar{q}q\rangle$,
$\langle g_{s}^{2}G^{2}\rangle=0.88\pm0.25~\mbox{GeV}^{4}$, $\langle\bar{q}g_{s}\sigma Gq\rangle=m_{0}^{2}\langle\bar{q}q\rangle$,
$\langle\bar{s}g_{s}\sigma Gs\rangle=m_{0}^{2}\langle\bar{s}s\rangle$
and $m_{0}^{2}=0.8\pm0.1\,\mbox{GeV}^{2}$ at the energy scale $\mu=1\ {\rm GeV}$.

Baryon masses used in the analysis of decay constants are given in Table \ref{Tab:doubly_heavy_mass}.
For the mass of $\Xi_{cc}^{++}$, we adopt the experimental value~\cite{Aaij:2017ueg},
and we use the isospin symmetry for the $\Xi_{cc}^{+}$. For other
baryons, we use the Lattice QCD results from Ref.~\cite{Brown:2014ena}.

%%%%%%%%%%%%%%%%%%%%%%%
\begin{table}[!htb]
\caption{Masses (in units of GeV) of doubly heavy baryons. We adopt the experimental
value for the mass of $\Xi_{cc}$~\cite{Aaij:2017ueg} and the Lattice
QCD results from Ref.~\cite{Brown:2014ena}.}
\label{Tab:doubly_heavy_mass} %
\begin{tabular}{c|c|c|c|c|c|c}
\hline 
Baryons  & $\Xi_{cc}$  & $\Omega_{cc}$  & $\Xi_{bb}$  & $\Omega_{bb}$  & $\Xi_{bc}$  & $\Omega_{bc}$ \tabularnewline
\hline 
Masses  & $3.621$ \cite{Aaij:2017ueg}  & $3.738$ \cite{Brown:2014ena}  & $10.143$ \cite{Brown:2014ena}  & $10.273$\cite{Brown:2014ena}  & $6.943$ \cite{Brown:2014ena}  & $6.998$ \cite{Brown:2014ena} \tabularnewline
\hline 
\end{tabular}
\end{table}

The continuum threshold $\sqrt{s_{0}}$ is used as $0.4\sim0.6$ GeV
higher than the corresponding baryon mass, where we have assumed that
the energy gap between the ground states and the first radial excited
states is approximately $0.5$ GeV~\cite{Wang:2012kw}.

Complying with the standard procedure of QCD sum rule analysis, the
Borel parameter $M^{2}$ is varied to find the optimal stability window,
in which the perturbative contribution should be larger than the condensate
contributions and meanwhile the pole contribution larger than the
continuum contribution.

The sum rule in Eq.~(\ref{eq:sum_rule_1}) will be numerically analyzed
since it is expected to have a better convergence in contrast with
the sum rule in Eq.~(\ref{eq:sum_rule_1_no_use}).

\subsection{Masses}

%%%%%%%%%%%%%%%%%%%%%%
\begin{figure}
\includegraphics[width=1\columnwidth]{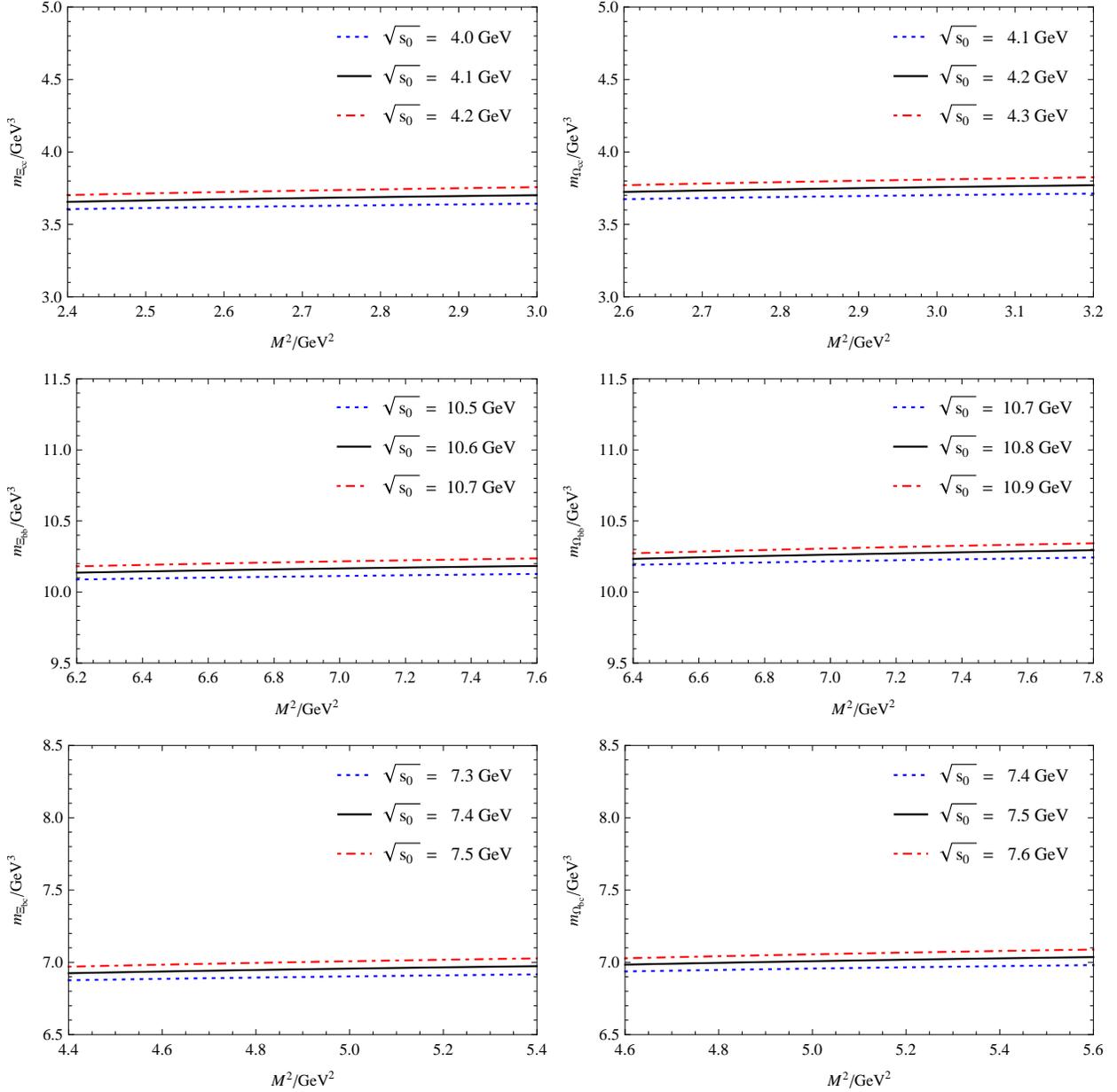}
\caption{The $M^{2}$-dependence of the masses of $\Xi_{cc}$, $\Omega_{cc}$
(the top two figures), $\Xi_{bb}$, $\Omega_{bb}$ (the middle two
figures), $\Xi_{bc}$ and $\Omega_{bc}$ (the bottom two figures).
The sum rule in Eq.~(\ref{eq:sum_rule_1}) is considered. The inputs
are taken as: $m_{c}=1.35\,{\rm GeV}$, $m_{b}=4.60\,{\rm GeV}$,
and $m_{s}=0.12\,{\rm GeV}$, condensate parameters are taken at $\mu=1\,{\rm GeV}$.}
\label{fig:mH} 
\end{figure}

%%%%%%%%%%%%%%%%%%%%%%
%%%%%%%%%%%%%%%%%%%%%%
\begin{figure}
\includegraphics[width=1\columnwidth]{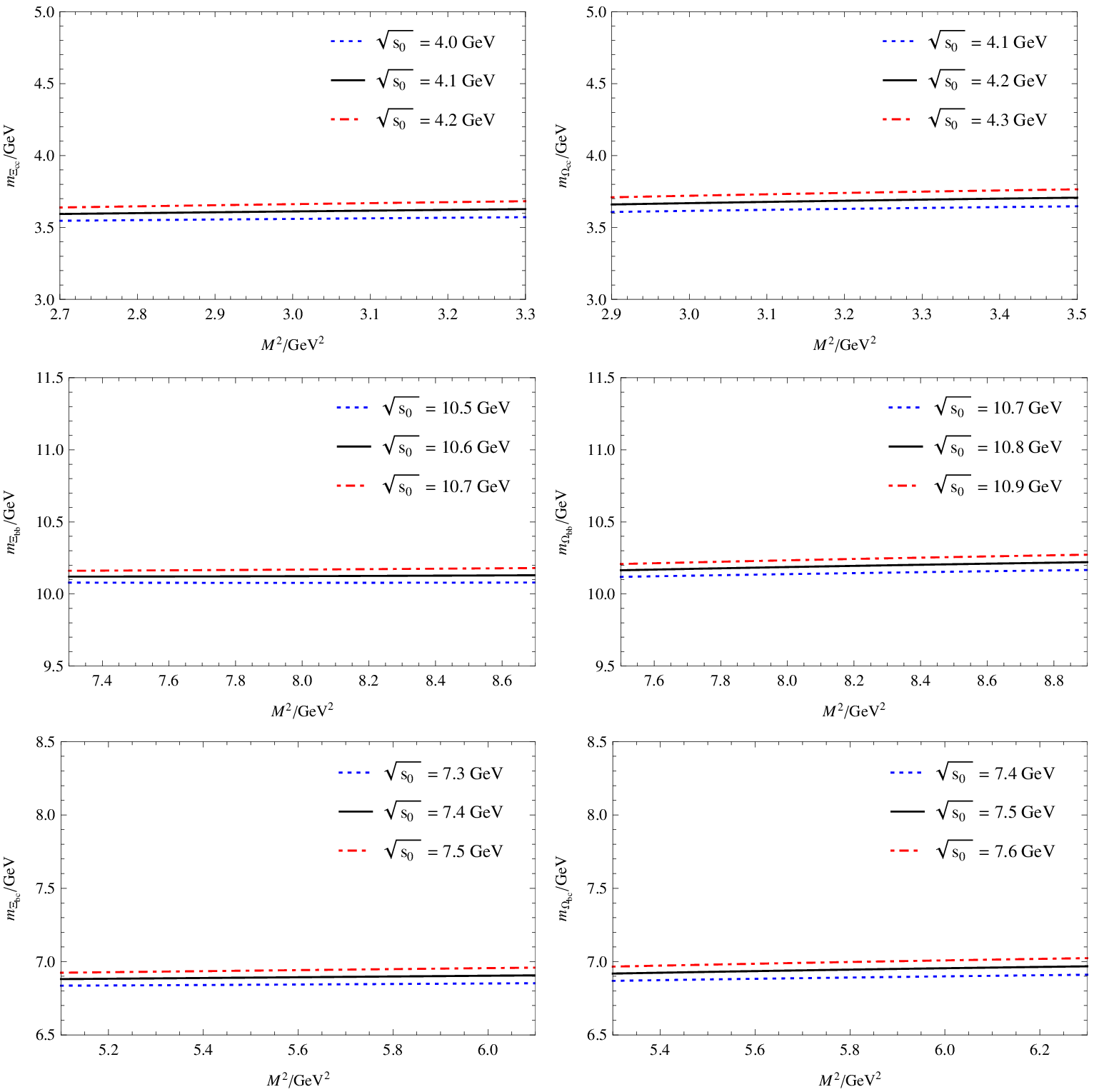}
\caption{Same as Fig.~\ref{fig:mH} but for the sum rule in Eq.~(\ref{eq:sum_rule_2})
is considered.}
\label{fig:mH_2} 
\end{figure}

%%%%%%%%%%%%%%%%%%%%%%

%%%%%%%%%%%%%%%%%%%%%%

%%%%%%%%%%%%%%%%%%%%%%
\begin{table}
\caption{Theoretical predictions for the masses (in units of ${\rm GeV}$)
of the doubly heavy baryons. The results of ``This work \#1'' are
predicted using Eq.~(\ref{eq:sum_rule_1}) while those of ``This
work \#2'' using Eq.~(\ref{eq:sum_rule_2}). The uncertainties of
the relavant parameters, including $M^{2}$, $s_{0}$, the quark masses
and the condensates, have been taken into account. For purposes of
comparison, some other QCDSR results from Ref. \cite{Zhang:2008rt}
and Ref. \cite{Wang:2010hs} and the Lattice QCD results from Ref.~\cite{Brown:2014ena}
are listed. Our results are consistent with Ref. \cite{Wang:2010hs}
and Ref.~\cite{Brown:2014ena} but somewhat different from Ref. \cite{Zhang:2008rt}.}
\label{Tab:mass} %
\begin{tabular}{c|c|c|c|c|c|c}
\hline 
Baryon  & This work \#1  & This work \#2 & Ref.~\cite{Zhang:2008rt}  & Ref. \cite{Wang:2010hs}  & Ref.~\cite{Brown:2014ena}  & Experiment\tabularnewline
\hline 
$\Xi_{cc}$  & $3.68\pm0.08$  & $3.61\pm0.09$ & $4.26\pm0.19$  & $3.57\pm0.14$  & $3.610\pm0.023\pm0.022$  & $3.6214\pm0.0008$\tabularnewline
\hline 
$\Omega_{cc}$  & $3.75\pm0.08$  & $3.69\pm0.09$ & $4.25\pm0.20$  & $3.71\pm0.14$  & $3.738\pm0.020\pm0.020$  & - -\tabularnewline
\hline 
$\Xi_{bb}$  & $10.16\pm0.09$  & $10.12\pm0.10$ & $9.78\pm0.07$  & $10.17\pm0.14$  & $10.143\pm0.030\pm0.023$  & - -\tabularnewline
\hline 
$\Omega_{bb}$  & $10.27\pm0.09$  & $10.19\pm0.10$ & $9.85\pm0.07$  & $10.32\pm0.14$  & $10.273\pm0.027\pm0.020$  & - -\tabularnewline
\hline 
$\Xi_{bc}$  & $6.95\pm0.09$  & $6.89\pm0.10$ & $6.75\pm0.05$  & - -  & $6.943\pm0.033\pm0.028$  & - -\tabularnewline
\hline 
$\Omega_{bc}$  & $7.01\pm0.09$  & $6.95\pm0.09$ & $7.02\pm0.08$  & - -  & $6.998\pm0.027\pm0.020$  & - -\tabularnewline
\hline 
\end{tabular}
\end{table}

%%%%%%%%%%%%%%%%%%%%%%

Differentiating Eq.~(\ref{eq:sum_rule_1}) or Eq.~(\ref{eq:sum_rule_2})
with respect to $-1/M^{2}$, one can extract the mass of the doubly
heavy baryon as
\begin{equation}
m_{H}^{2}=\frac{\int_{\Delta}^{s_{0}}ds\rho_{1}(s)se^{-s/M^{2}}}{\int_{\Delta}^{s_{0}}ds\rho_{1}(s)e^{-s/M^{2}}}
\end{equation}
or
\begin{equation}
m_{H}^{2}=\frac{\frac{d}{d(-1/M^{2})}\int_{\Delta}^{s_{0}}ds\left[\sqrt{s}\rho_{1}(s)+\rho_{2}(s)\right]e^{-s/M^{2}}}{\int_{\Delta}^{s_{0}}ds\left[\sqrt{s}\rho_{1}(s)+\rho_{2}(s)\right]e^{-s/M^{2}}}.
\end{equation}

The optimal stability window for $M^{2}$ can be determined as follows.
The upper bounds of the Borel parameters $M^{2}$ can be determined
by the requirement that the pole contribution should be larger than
the continuum contribution, while the lower bound can be determined
by the requirement that the perturbative contribution should be larger
than the quark condensate contribution. For the sum rule in Eq.~(\ref{eq:sum_rule_2}),
$M_{{\rm max}}^{2}=3.3,\ 3.5,\ 8.7,\ 9.5,\ 6.1,\ 6.3\ {\rm GeV}^{2}$
and $M_{{\rm min}}^{2}=2.7,\ 2.0,\ 7.1,\ 5.2,\ 4.7,\ 3.6\ {\rm GeV}^{2}$
for $\Xi_{cc}$, $\Omega_{cc}$, $\Xi_{bb}$, $\Omega_{bb}$, $\Xi_{bc}$
and $\Omega_{bc}$ respectively. The optimal windows for $M^{2}$
can be chosen as $[2.7,3.3],\ [2.9,3.5],\ [7.3,8.7],\ [7.5,8.9],\ [5.1,6.1]$
and $[5.3,6.3]$ respectively. For the sum rule in Eq.~(\ref{eq:sum_rule_1}),  the optimal windows for $M^{2}$
can be chosen as $[2.4,3.0],\ [2.6,3.2],\ [6.2,7.6],\ [6.4,7.8],\ [4.4,5.4]$
and $[4.6,5.6]$ respectively.

Dependence of the predicted mass $m_{H}$ on the Borel parameter $M^{2}$
is shown in Figs.~\ref{fig:mH} and \ref{fig:mH_2}, where the sum
rules in Eq.~(\ref{eq:sum_rule_1}) and Eq.~(\ref{eq:sum_rule_2})
are adopted, respectively. Using Eq.~(\ref{eq:sum_rule_1}),  we obtain
\[
m_{\Xi_{cc}}=(3.68\pm0.08)\ {\rm GeV},
\]
where only the positive parity baryons are taken into account. When the contamination from the $1/2^-$ baryon is considered, we find the mass is slightly changed
\[
m_{\Xi_{cc}}=(3.61\pm0.09)\ {\rm GeV}. 
\] 
Here the uncertainties of the relevant parameters, including
$M^{2}$, $s_{0}$, the quark masses and the condensates, have been
taken into account. It can be seen that our values are consistent
with the experimental value when the errors are taken into account.
Our results are also consistent with other estimates for instance
Ref.~\cite{Wang:2010hs}. A collection of the results can be found
in Table \ref{Tab:mass}.

\subsection{Decay constants}

\begin{figure}
\includegraphics[width=1\columnwidth]{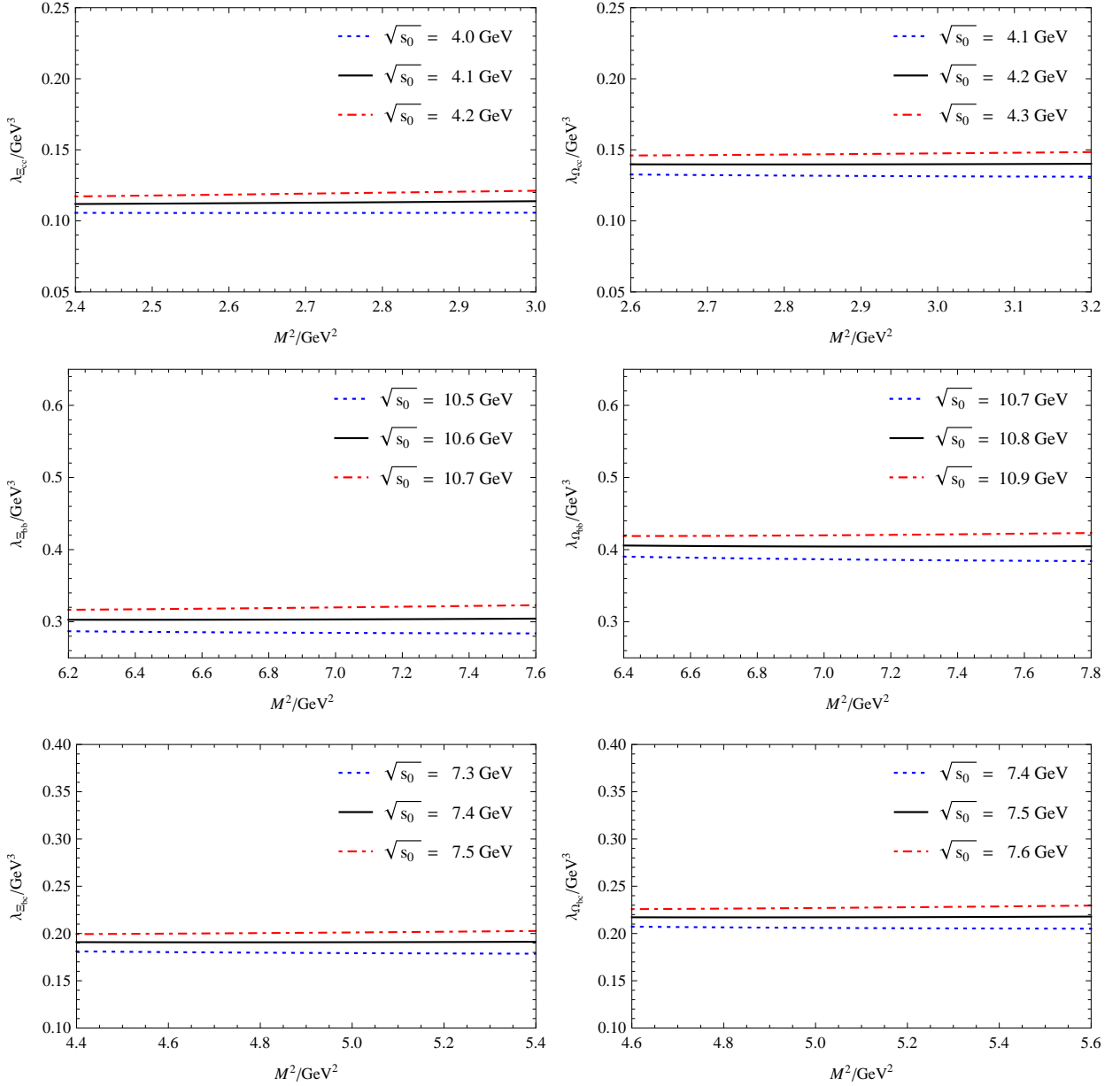}
\caption{The $M^{2}$-dependence of the decay constants of $\Xi_{cc}$,
$\Omega_{cc}$ (the top two figures), $\Xi_{bb}$, $\Omega_{bb}$ (the
middle two figures), $\Xi_{bc}$ and $\Omega_{bc}$ (the bottom two
figures). The continuum threshold are taken as $\sqrt{s_{0}}=4.0\sim4.2$ GeV, $\sqrt{s_{0}}=4.1\sim4.3$
GeV, $\sqrt{s_{0}}=10.5\sim10.7$ GeV, $\sqrt{s_{0}}=10.7\sim10.9$
GeV, $\sqrt{s_{0}}=7.3\sim7.5$ GeV and $\sqrt{s_{0}}=7.4\sim7.6$
GeV, respectively. The sum rule in Eq.~(\ref{eq:sum_rule_1}) is
considered.}
\label{fig:decay_constant} 
\end{figure}

\begin{figure}
\includegraphics[width=1\columnwidth]{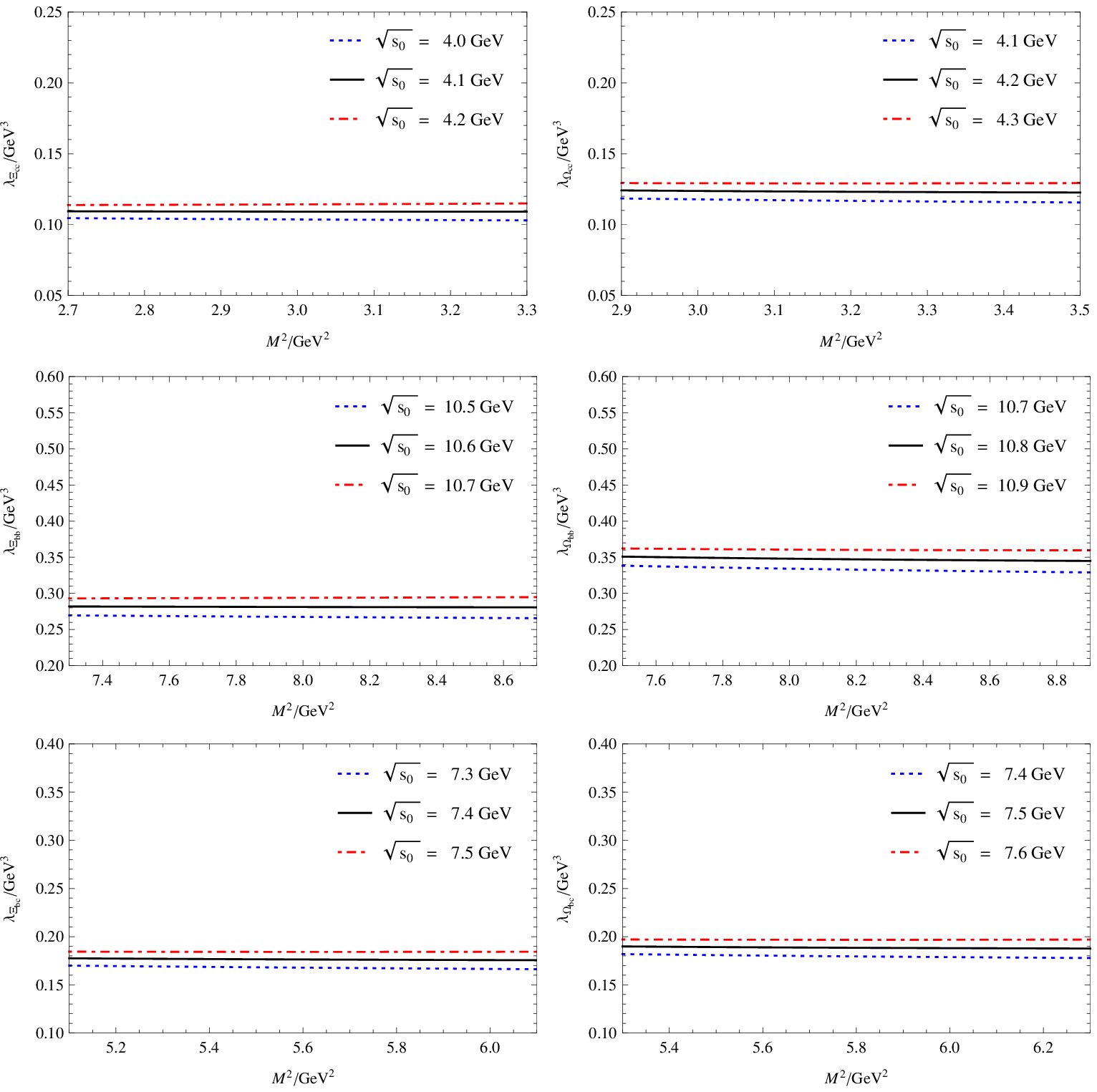}
\caption{Same as Fig.~\ref{fig:decay_constant} but for the sum rule in Eq.~(\ref{eq:sum_rule_2})
is considered. }
\label{fig:decay_constant_2} 
\end{figure}

Dependence of the ``decay constants\char`\"{} $\lambda_{H}$ on the
Borel parameter $M^{2}$ is shown in Figs.~\ref{fig:decay_constant}
and \ref{fig:decay_constant_2}, where the sum rules in Eq.~(\ref{eq:sum_rule_1})
and Eq.~(\ref{eq:sum_rule_2}) are adopted, respectively. Numerical
results for the ``decay constants\char`\"{} can be found in Table
\ref{Tab:decay_constants}.

%%%%%%%%%%%%%%%%%%%%%%

%%%%%%%%%%%%%%%%%%%%%%
%%%%%%%%%%%%%%%%%%%%%%
%%%%%%%%%%%%%%%%%%%%%%
%%%%%%%%%%%%%%%%%%%%%%

%%%%%%%%%%%%%%%%%%%%%%

%%%%%%%%%%%%%%%%%%%%%%
%%%%%%%%%%%%%%%%%%%%%%
\begin{table}
\caption{Decay constants $\lambda_{H}$ (in units of ${\rm GeV}^{3}$) for
the doubly heavy baryons. The results of ``This work \#1'' are predicted
using Eq.~(\ref{eq:sum_rule_1}) while those of ``This work \#2''
using Eq.~(\ref{eq:sum_rule_2}). The uncertainties of the relavant
parameters, including $M^{2}$, $s_{0}$, the quark masses, the condensates
and the baryon masses, have been taken into account. For purposes
of comparison, the QCDSR results from Ref. \cite{Wang:2010hs} are
listed.}
\label{Tab:decay_constants} %
\begin{tabular}{c|c|c|c}
\hline 
Baryon  & This work \#1 & This work \#2 & Ref. \cite{Wang:2010hs} \tabularnewline
\hline 
$\Xi_{cc}$  & $0.113\pm0.029$  & $0.109\pm0.021$  & $0.115\pm0.027$ \tabularnewline
\hline 
$\Omega_{cc}$  & $0.140\pm0.033$  & $0.123\pm0.024$  & $0.138\pm0.030$ \tabularnewline
\hline 
$\Xi_{bb}$  & $0.303\pm0.094$  & $0.281\pm0.071$  & $0.252\pm0.064$ \tabularnewline
\hline 
$\Omega_{bb}$  & $0.404\pm0.112$  & $0.347\pm0.083$  & $0.311\pm0.077$ \tabularnewline
\hline 
$\Xi_{bc}$  & $0.191\pm0.053$  & $0.176\pm0.040$  & - -\tabularnewline
\hline 
$\Omega_{bc}$  & $0.217\pm0.056$  & $0.188\pm0.041$  & - -\tabularnewline
\hline 
\end{tabular}
\end{table}

%%%%%%%%%%%%%%%%%%%%%%
%%%%%%%%%%%%%%%%%%%%%%

A few remarks are given in order. 
\begin{itemize}
\item It is necessary to point out that when including the contributions
from the $1/2^{-}$ baryons the threshold parameter might be somewhat
higher. In this analysis, we have approximately use the same values. 
\item Comparing the two sets of results in Table~\ref{Tab:decay_constants},
one can see that the negative parity baryons do not provide significant
modifications.
\item We can see from Table~\ref{Tab:decay_constants} that the decay constants
of $\Omega_{QQ^{\prime}}$ are slightly larger than those of $\Xi_{QQ^{\prime}}$. 
\end{itemize}

\section{Conclusion}

In this work we have calculated the ``decay constants" for doubly heavy
baryons $\Xi_{cc}$, $\Omega_{cc}$, $\Xi_{bb}$, $\Omega_{bb}$,
$\Xi_{bc}$ and $\Omega_{bc}$ using  QCD sum rules.
In the calculation we have included both  the positive and negative
parity baryons, and found that the $1/2^-$ contamination is not severe.  The extracted results for the decay constants are ingredients for
the study of weak decays and other properties of doubly heavy baryons, including the lifetimes~\cite{Colangelo:1996ta,Lenz:2014jha,Kirk:2017juj}.

\section*{Acknowledgements}

The authors are grateful to Pietro Colangelo, Fulvia De Fazio,    Zhi-Gang Wang, Yu-Ming Wang, and Fu-Sheng Yu  for useful discussions. This work
is supported in part by National Natural Science Foundation of China
under Grant No.11575110, 11655002, 11735010, Natural Science Foundation of Shanghai
under Grant No.~15DZ2272100 and No.~15ZR1423100,  by Shanghai Key Laboratory for Particle Physics and Cosmology,  and by Key Laboratory
for Particle Physics, Astrophysics and Cosmology, Ministry of Education.

%\appendix

%\cite{Mattson:2002vu}

\end{document}